\documentclass{article}

\title{Integral Transforms in Relativistic Quantum Constraint Mechanics}
\author{Robert J. Ducharme}

\begin{document}
\maketitle

\centerline{151 Fairhills Dr., Ypsilanti, MI 48197}
\centerline{E-mail: ducharme01@comcast.net}

\begin{abstract}
In relativistic quantum constraint mechanics the state of a physical system is constrained to a 3-dimensional subspace of Minkowski 4-space. Fourier transformation can be used to relate this state between constraint spaces in 4-position and 4-momentum space. It is shown that integral transforms of this nature can be carried out using Lorentz-invariant 3-dimensional constraint space coordinates such that a complete equivalence class of 4-space representations can be constructed from the transform. This method is further applied to develop a relativistic generalization of the Segal-Bargmann transformation that leads to the representation of quantum systems in a three-dimensional subspace of Bargmann 4-space.
\end{abstract}

\section{Introduction}
Quantum Constraint Mechanics (QCM) is a generalized form of Hamiltonian
quantum mechanics \cite{AK, RST, CA} that enables bound systems of interacting particles to be described in a manifestly covariant manner. QCM has been applied to problems in nuclear \cite{LC,WC} and high-energy physics \cite{CWA}. It assumes each particle in a system of spinless particles can be described using a single K-G equation that includes an interaction potential to take account of the influence of the other particles. The approach is to seek solutions of these equations that correspond to solutions of the Schr\"{o}dinger equation for the system in the non-relativistic limit. It is significant that QCM evades the Currie-Jordan-Sudershan no-interaction theorem \cite{CJS} thus distinguishing it from the Bakamjiam-Thomas method \cite{BDK}.

The purpose of this paper is to present Lorentz-invariant integral transforms for relating state functions between different 4-space representations in quantum constraint mechanics. In pursuit of this goal it has been found convenient to introduce a system of constraint space coordinates $\xi_i (i=1,2,3)$ measured with respect to an orthogonal set of axes that are also inside the constraint space. The concept of expressing a quantum constraint mechanics problem in constraint space coordinates might be compared to using planar coordinates in a plane. In both cases, the coordinate systems are internal thus decoupling them from the orientation of the subspace in the higher-dimensional space that contains it. Thus, a unique state function $\psi(\xi_i)$ in a constraint space maps into an equivalence class $\psi(x_\perp^\mu) (\mu = 1,2,3,4)$ of representations in the 4-position space $x_\mu$ but $\psi(\xi_i)$ is usually simpler to integrate over the constraint space since there is no need to impose an artificial relationship between the components of $x_\mu$. 

Constraint space coordinates $\xi_i(x_\mu)$ for a 4-position space $x_\mu$ and $\pi_i(p_\mu)$ for a 4-momentum space $p_\mu$ are introduced in section 2. It is shown that these constraint space coordinates form Lorentz-invariant 3-vectors such that their components are independent of each other and their magnitudes are the same for all observers. It is also shown that $\xi_i$ and $\pi_i$ respectively correspond to 3-position $x_i$ and 3-momentum $p_i$ in the non-relativistic limit. 

The quantum constraint equations for a relativistic quantum bound system of two particles interacting through a scalar potential are presented in section 3 of this paper. These consist of a free-field K-G equation that describes the state of the system in center-of-mass coordinates alongside two internal constraint equations that together determine the quantum state of the particles inside the constraint space. It is shown that these two internal constraint equation can be reduced to a single three-dimensional equation in terms of $\xi_i$-coordinates. The normalization of the $\psi(\xi_i)$ state function and Fourier transformation between the $\xi_i$ and $\pi_i$ constraint spaces are also discussed.

In section 4 the solution of the quantum constraint equations in constraint space coordinates is obtained for the relativistic three-dimensional harmonic oscillator. The fully relativistic form of the ladder operators for raising and lowering the eigenstates of the oscillator are also determined as well as the Lorentz invariant form of the Segal-Bargmann transformation giving the oscillator functions in Bargmann 4-space.

The 4-vector convention in this paper will be $x_\mu = (x_1, x_2, x_3, x_4)$ and $x_\mu =(x_1, x_2, x_3, -x_4)$. Natural units, $c = \hbar = 1$ will be used throughout.

\section{Constraint Space Coordinates}
Consider a quantum bound system consisting of two interacting particles each to be represented using an index $k(=1,2)$.  Let $m_{k}$, $x_{k\mu}$ and $p_{k\mu}$ denote the mass, 4-position and 4-momentum of each of the particles respectively. The center-of-mass and relative coordinates \cite{CA} for the system can then be expressed as
\begin{equation} \label{eq: xdef}
X_{\mu} = \eta_1 x_{1\mu} + \eta_2 x_{2\mu}, \quad x_{\mu} = x_{1\mu} - x_{2\mu}
\end{equation}
in 4-position space
\begin{equation} \label{eq: pdef}
P_{\mu} = p_{1\mu} + p_{2\mu}, \quad p_{\mu} = \eta_2 p_{1\mu} - \eta_1 p_{2\mu}
\end{equation}
in 4-momentum space. Here the $\eta_1$ and $\eta_2$ parameters take the form
\begin{equation} \label{eq: epsilon1}
\eta_1 = \frac{1}{2}+\frac{m_1^2-m_2^2}{M_0^2}, \quad \eta_2 = \frac{1}{2}-\frac{m_1^2-m_2^2}{M_0^2}
\end{equation}
It will be further assumed that the oscillator is in a free state such that the total momentum $P_{\mu}$ satisfies the relationship
\begin{equation} \label{eq: totMom}
P^\mu P_\mu = -M_0^2
\end{equation}
where $M_0$ is the total mass of the oscillator. 

It is usual in QCM, to further define a set of 4-space coordinates that locate position coordinates inside the constraint space. These coordinates are defined for both 4-position space
\begin{equation} \label{eq: xperp}
x_{\perp}^\mu = x^\mu + M_0^{-2}P^\mu (P_\nu x^\nu)
\end{equation}
and 4-momentum space
\begin{equation} \label{eq: pperp}
p_{\perp}^\mu = p^\mu + M_0^{-2} P^\mu (P_\nu p^\nu)
\end{equation}
Pre-multiplying these results by $P_\mu$ gives
\begin{equation} \label{eq: xcon}
P_\mu x_{\perp}^\mu = 0
\end{equation}
\begin{equation} \label{eq: pcon}
P_\mu p_{\perp}^\mu = 0
\end{equation}
showing $x_{\perp}^\mu = (x^1,x^2,x^3,0)$ and $p_{\perp}^\mu = (p^1,p^2,p^3,0)$ in the rest frame of the bound system. 

For present purposes, it is convenient to define a set of constraint space coordinates $\xi_i$ referred to axes that all fall inside the constraint space. It is clear
\begin{equation} \label{eq: xi_rest}
\xi_i = x_i^\prime
\end{equation}
in the rest frame of the bound system but the requirement is for a representation that is valid in all inertial frames of reference. This can be obtained through the general Lorentz transformation
\begin{equation} \label{eq: glt}
x_i^{\prime} = x_i + \gamma v_i \left(\frac{\gamma v_j x_j }{1+\gamma} - x_4 \right)
\end{equation}
\begin{equation}
x_4^{\prime} = \gamma(x_4-v_i x_i)
\end{equation}
(see ref. \cite{SR}). Here, $v_i$ is the velocity of the system and $\gamma = (1-v^2)^{-1/2}$. Thus, inserting eq. (\ref{eq: glt}) in eq. (\ref{eq: xi_rest}) gives
\begin{equation} \label{eq: xi_vx}
\xi_i = x_i + \gamma v_i \left(\frac{\gamma v_j x_j }{1+\gamma} - x_4 \right)
\end{equation}
or equivalently
\begin{equation} \label{eq: xi_px}
\xi_i = x_i + \frac{P_i(P_\mu x^\mu - M_0 x_4)}{M_0(M_0+P_4)} 
\end{equation}
where eq. (\ref{eq: xi_rest}) is recovered in the rest frame of the bound system. The differential form of eq. (\ref{eq: xi_px}):
\begin{equation} \label{eq: dxi_px}
\frac{\partial}{\partial \xi_i} = \frac{\partial}{\partial x_i} + \frac{P_i}{M_0+P_4}\left( \frac{P_\mu}{M_0} \frac{\partial}{\partial x_\mu} - \frac{\partial}{\partial x_4} \right)
\end{equation}
follows from the chain rule for partial differentiation.

The 3-vector $\xi_i$ is readily shown to have the following Lorentz invariant  properties:
\begin{equation} \label{eq: xi_squared}
\xi^2 = x_\perp^2 = x^2+M_0^{-2}(P_\mu x^\mu)^2
\end{equation}
\begin{equation} \label{eq: xi_independ}
\frac{\partial \xi_j}{\partial \xi_i}=\delta_{ij}
\end{equation}
where $\delta_{ij}$ is the Kronecker delta. These results confirm that the 3-vector $\xi_i$ has three orthogonal components in all inertial frames of reference and a Lorentz-invariant length. The correspondence of $\xi_i$ to $x_i$ in the rest frame of the system suggests replacement rule:
\begin{equation} \label{eq: subRule_x}  
x_i \rightarrow \xi_i
\end{equation}
as a means of taking certain non-relativistic equations over to a Lorentz-invariant form. 

The foregoing arguments for the constraint space $\xi_i$ in 4-position space can also be made to develop the idea of a constraint space  $\pi_i$ in 4-momentum space. In the 4-momentum case, the $\pi_i=p_i$ correspondence requirement in the rest frame of the system leads to the following Lorentz-invariant definition 
\begin{equation} \label{eq: pi_pp}
\pi_i = p_i + \frac{P_i(P_\mu p^\mu - M_0 p_4)}{M_0(M_0+P_4)} 
\end{equation}
analogous to the derivation of eq. (\ref{eq: xi_px}). The components of $\pi_i$ are orthogonal in all inertial reference frames and the magnitude of the $\pi_i$ can be determined from the Lorentz covariant expression
\begin{equation} \label{eq: pi_squared}
\pi^2 = p_\perp^2 = p^2+M_0^{-2}(P_\mu p^\mu)^2
\end{equation}
similar to eq. (\ref{eq: xi_squared}). The replacement rule in $\pi_i$ constraint space is
\begin{equation} \label{eq: subRule_p}  
p_i \rightarrow \pi_i
\end{equation}
similar to eq. (\ref{eq: subRule_x} ).

The quantum mechanical operators for position and momentum are
\begin{equation} \label{eq: op_x}
\hat{\xi}_i = \xi_i \quad \hat{\pi}_i = -\imath \frac{\partial}{\partial \xi_i}
\end{equation}
in the position constraint space and
\begin{equation} \label{eq: op_p}
\hat{\xi}_i = -\imath \frac{\partial}{\partial \pi_i} \quad \hat{\pi}_i = \pi_i 
\end{equation}
in the momentum constraint space. These results follow from application of the substitution rules (\ref{eq: subRule_x}) and (\ref{eq: subRule_p}) to the non-relativistic operators in 3-position and 3-momentum space.

\section{Quantum Constraint Mechanics}
The pair of coupled K-G equations constraining the wave function $\psi$ for two particles bound together through a scalar function $\phi(x_\perp^\mu)$ can be expressed as
\begin{equation} \label{eq: KG_BOUND}
(\hat{p}_k^2+m_k^2+\Phi)\psi=0
\end{equation}
Use of eqs. (\ref{eq: xdef}) and (\ref{eq: totMom}) enable eq. (\ref{eq: KG_BOUND}) to be transformed into center-of-mass and relative coordinates giving
\begin{equation} \label{eq: KG_COM}
\frac{\partial^2 \psi}{\partial X^2} = M_0^2\psi
\end{equation}
\begin{equation} \label{eq: KG_INTER}
\frac{\partial^2 \phi}{\partial x^2} + 2(\sigma-\Phi)\phi = 0
\end{equation}
\begin{equation} \label{eq: KG_SCON}
P_\mu \frac{\partial \phi}{\partial x_\mu} = 0
\end{equation}
having assumed a separable solution of the form
\begin{equation} \label{eq: SEP_SOL}
\psi = \phi(x_\perp^\mu)\exp(\imath P_\mu X^\mu)
\end{equation}
Here the rest mass of the bound system can be determined from the condition
\begin{equation} \label{eq: KG_REST_MASS}
M_0 = \sqrt{m_1^2+m_2^2+4\sigma+\sqrt{[(m_1^2+m_2^2+4\sigma)^2+(m_1^2-m_2^2)^2]}}
\end{equation}
and $\sigma$ is the constant of separation. It can be seen that eq. (\ref{eq: KG_REST_MASS}) gives $M_0 =m_1+m_2$ for the free-particle case $(\sigma = 0)$. The positive value of the square root is indicated since the intention is to seek relativistic solutions that correspond to solutions of the Schr\"{o}dinger equation in the non-relativistic limit. It is shown next that eqs. (\ref{eq: KG_INTER}) and (\ref{eq: KG_SCON}) do reduce to the Schr\"{o}dinger equation in the non-relativistic limit.

For $\sigma \ll m_1$ and $\sigma \ll m_2$, the rest mass $M_0$ can be approximated to the non-relativistic form
\begin{equation}
M_0 \simeq m_1 + m_2 + E 
\end{equation}
where $E$ is the non-relativistic energy of the system
\begin{equation} \label{eq: energy_nr}
E=\frac{\sigma}{m_r}
\end{equation}
and
\begin{equation}
m_r = \frac{m_1 m_2}{m_1+m_2}
\end{equation}
is the reduced mass of the particles. If the velocity of the bound system is also
much smaller than the speed of light, eq. (\ref{eq: KG_SCON}) implies $\partial \phi / \partial x_4 \simeq 0$ showing $\psi$ is independent of the relative time $x_4$. This result alongside eq. (\ref{eq: energy_nr}) enable eqs. (\ref{eq: KG_INTER}) and (\ref{eq: KG_SCON}) to be approximated to the Schr\"{o}dinger equation:
\begin{equation}
\frac{\hbar^2}{2m_r}\frac{\partial^2 \phi}{\partial x_i^2}+V\phi=E\phi
\end{equation}
having restored $\hbar$ and put $V = \sigma / 2m_r$.

The above argument shows that eqs. (\ref{eq: KG_INTER}) and (\ref{eq: KG_SCON}) can be combined in the non-relativistic limit to give a single equation for $\psi$ in three-dimensions. The process of reducing eqs. (\ref{eq: KG_INTER}) and (\ref{eq: KG_SCON}) to a single three-dimensional equation is also possible in relativistic frames with the help of eq. (\ref{eq: dxi_px}). The first step is to show
\begin{equation} \label{eq: dxi_squared}
\frac{\partial^2}{\partial \xi_i^2} = \frac{\partial^2}{\partial x_\mu^2}-\left(\frac{P_\mu}{M_0}\frac{\partial}{\partial x_\mu}\right)^2
\end{equation}
Eqs. (\ref{eq: KG_INTER}) can then be written as
\begin{equation} \label{eq: KG_XI}
\frac{\partial^2\phi}{\partial \xi_i^2} + 2(\sigma-\Phi)\phi = 0
\end{equation} 
having used eq. (\ref{eq: KG_SCON}). Here, eq, (\ref{eq: xi_squared}) enables relativistic potentials of the form $\Phi(x_\perp^\mu)$ to be easily transformed into $\xi_i$-coordinates. The normalization condition for $\psi(\xi_i)$ is
\begin{equation} \label{eq: normalCond}
\int \psi^* \psi d^3\xi=1
\end{equation}
where $d^3\xi = d\xi_1d\xi_2d\xi_3$ denotes a volume element in the rectangular coordinates system of the constraint space. Integrating over the constraint space in $\xi_i$ coordinates is generally more convenient than integrating over the constraint space in $x_\mu$-coordinates. Specifically, a volume element of the constraint space in $x_\mu$-coordinates must be written $d^3x_\perp = C(x_\mu)dx_1dx_2dx_3$ where $C(x_\mu)$ is an arbitrary constraint condition chosen for the purpose of reducing the number of independent variables from four to three. 

The Fourier transformation relating a function $f$ between the position and momentum constraint spaces is
\begin{equation} \label{eq: ftrans}
f[\pi_i(p_\mu)] = \frac{1}{\sqrt{8\pi^3}} \int g[\xi_i(x_\mu)] \exp(-\imath \pi_i \xi_i) d^3\xi
\end{equation} 
with the inverse form
\begin{equation} \label{eq: ftrans_inv}
g[\xi_i(x_\mu)] = \frac{1}{\sqrt{8\pi^3}}\int f[\pi_i(p_\mu)] \exp(\imath \pi_i \xi_i) d^3\pi
\end{equation}
Here, there is an advantage to avoiding the introduction of the $C(x_\mu)$ constraint condition since $g(\xi_i)$ has an equivalence class of representations in $x_\mu$-coordinates that the $C(x_\mu)$ condition restricts. Integral transforms based on constraint coordinates are therefore more general than methods that fix the constraint space in terms of 4-space coordinates.

\section{The Harmonic Oscillator}
The relativistic form of the internal quantum constraint eq. (\ref{eq: KG_XI}) for the three-dimensional harmonic oscillator is
\begin{equation} \label{eq: KG_XI_HO}
-\frac{\partial^2\psi}{\partial \xi^2} + \Omega^2\xi^2\psi = 2\sigma \psi
\end{equation} 
where $\Omega$ is the spring constant of the oscillator. This gives the solution:
\begin{equation} \label{eq: psi1} 
\psi[\xi_i(x_\mu), X_\mu, P_\mu] = \phi_{l_1}(\xi_1)\phi_{l_2}(\xi_2)\phi_{l_3}(\xi_3)\exp(\imath P_\mu X^{\mu})
\end{equation}
(see ref. \cite{DFL}) where 
\begin{equation}
\phi_{l_i}(\xi_i) = \frac{(\Omega / \pi)^{1/4}}{\sqrt{2^{l_i} l_i!}}  H_{l_i}(\sqrt{\Omega}\xi_i) \exp \left( -\frac{\Omega \xi_i^2}{2} \right)
\end{equation}
$H_{l_j}$ are Hermite polynomials and $l_1,l_2,l_3$ are positive integers. The corresponding energy eigenvalues are
\begin{equation} \label{eq: energy} 
\sigma_n = \Omega \left(\frac{3}{2} + n \right)
\end{equation}
where $n=l_1+l_2+l_3$. Here, $\psi$ has been normalized using eq. (\ref{eq: normalCond}).

The relativistic ladder operator definitions
\begin{equation} \label{KG_OP_A}
\hat{\alpha}_i^\pm = \frac{1}{\sqrt{2 \Omega}} \left( \mp \frac{\partial}{\partial \xi_i} + \Omega \xi_i \right)
\end{equation}
enable eq. (\ref{eq: KG_XI_HO}) to be expressed in the concise form
\begin{equation} \label{KG_HO_A}
\left(\hat{\alpha}_i^+ \hat{\alpha}_i^- + \frac{3}{2} \right)\Omega \psi = \sigma_n \psi
\end{equation}
It is readily shown this equation reduces to
\begin{equation} \label{SCHROD_HO_A}
\left(\hat{a}_i^+ \hat{a}_i^- + \frac{3}{2} \right)\hbar \omega \psi = E_n \psi
\end{equation}
where
\begin{equation} \label{SCHROD_OP_A}
\hat{\alpha}_i^\pm \simeq \hat{a}_i^\pm =  \frac{1}{\sqrt{2 m_r \omega}} \left( \mp \frac{\partial}{\partial x_i} + m_r \omega x_i \right)
\end{equation}
in the non-relativistic limit having made use of eq. (\ref{eq: energy_nr}), put $\Omega = m_r \omega$ and restored $\hbar$. This is the familiar ladder operator form of the Schr\"{o}dinger equation for the harmonic oscillator.

The explicit form of eq. (\ref{KG_OP_A}) in terms of 4-space coordinates can be obtained with the help of eqs. (\ref{eq: xi_px}) and (\ref{eq: dxi_px}) to be
\begin{equation} \label{eq: explicitladders}
\hat{\alpha}_i^{\pm} = \frac{1}{\sqrt{2 \Omega}}\left[ \mp \left( \frac{\partial}{\partial x_i} - \frac{P_i}{M_0+P_4} \frac{\partial}{\partial x_4} \right) +  \Omega x_i + \Omega\frac{P_i}{M_0} \left( \frac{P_jx_j}{M_0+P_4}-x_4\right) \right]
\end{equation} 
Thus, applying eqs. (\ref{KG_OP_A}) or (\ref{eq: explicitladders}) to the oscillator function (\ref{eq: psi1}) it can be shown
\begin{equation} \label{eq: l0}
\hat{\alpha}_i^{-} \phi_{l_i}[\xi_i(x_\mu)] = \sqrt{l_i} \phi_{l_i-1}[\xi_i(x_\mu)]
\end{equation}
lowers the quantum state of the oscillator, and
\begin{equation} \label{eq: r0}
\hat{\alpha}_i^{+} \phi_{l_i}[\xi_i(x_\mu)] = \sqrt{l_i+1} \phi_{l_i+1}[\xi_i(x_\mu)]
\end{equation} 
raises it. These fully relativistic expressions hold true in all inertial frames of reference.

Fourier and Segal-Bargmann transforms can now be calculated for the relativistic three-dimensional harmonic oscillator wave functions (\ref{eq: psi1}). In particular, eq. (\ref{eq: ftrans}) gives the Fourier transform to be
\begin{equation} \label{eq: psi2} 
\psi[\pi_i(p_\mu), X_\mu, P_\mu] = \phi_{l_1}(\pi_1)\phi_{l_2}(\pi_2)\phi_{l_3}(\pi_3)\exp(\imath P_\mu X^{\mu})
\end{equation}
where 
\begin{equation}
\phi_{l_i}(\pi_i) = \frac{(1 / \Omega \pi)^{1/4}}{\sqrt{2^{l_i} l_i!}}  H_{l_i}(\pi_i / \sqrt{\Omega}) \exp \left( -\frac{\pi_i^2}{2\Omega} \right)
\end{equation}
These are the relativistic harmonic oscillator wave functions in 4-momentum space.

In the current notation, the Segal-Bargmann transformation \cite{RNA} takes the form
\begin{equation} \label{sb_trans}
f(\alpha_i) = \left(\frac{\Omega}{\pi}\right)^{1/4}\int g(\xi_i) \exp \left(\frac{-\Omega \xi_i^2-\alpha_i^2}{2} \right) \exp \left( -\sqrt{2\Omega} \alpha_i \xi_i \right) d\xi_i
\end{equation}
having applied the replacement rule (\ref{eq: subRule_x}) to the non-relativistic form of this integral transform. Using eq. (\ref{sb_trans}) to transform the harmonic oscillation wave function (\ref{eq: psi1}) into the Bargmann 4-space $a_\mu$ gives
\begin{equation} \label{eq: psi3} 
\psi[\alpha_i(a_\mu), X_\mu, P_\mu] = \frac{1}{\sqrt{l_1!l_2!l_3!}} \alpha_1^{l_1}(a_\mu)\alpha_2^{l_2}(a_\mu)\alpha_3^{l_3}(a_\mu)\exp(\imath P_\mu X^{\mu})
\end{equation}
where 
\begin{equation} \label{eq: alpha_px}
\alpha_i = a_i + \frac{P_i(P_\mu a^\mu - M_0 a_4)}{M_0(M_0+P_4)} 
\end{equation}
Bargmann 4-space \cite{Low} has a clear connection to the 4-dimensional harmonic oscillator \cite{Kim} equation:
\begin{equation} \label{KG_HO_4D}
\left(\hat{a}_\mu^+ \hat{a}^{-\mu} + 1 \right)\Omega \psi = \sigma_n \psi
\end{equation}
where
\begin{equation} \label{KG_OP_4D}
\hat{a}_\mu^\pm = \frac{1}{\sqrt{2 \Omega}} \left( \mp \frac{\partial}{\partial x_\mu} + \Omega x_\mu \right)
\end{equation}
are the 4-dimensional ladder operators. It is therefore of interest to note that eq. (\ref{eq: alpha_px}) can also be written in the operator form
\begin{equation} \label{eq: op_alpha_px}
\hat{\alpha}_i = \hat{a}_i + \frac{P_i(P_\mu \hat{a}^\mu - M_0 \hat{a}_4)}{M_0(M_0+P_4)} 
\end{equation}
showing that the relativistic 3-dimensional ladder operators of QCM can be expressed as a linear combination of 4-dimensional ladder operators \cite{RJD}.

\section{Conclusion}
It is concluded that a wavefunction $\psi(x_\perp^\mu)$ in quantum constraint mechanics can, equivalently, be written in the form $\psi[\xi_i(x_\mu)]$ where $\xi_i$ is a Lorentz-invariant 3-vector. The components of $\xi_i$ have been interpreted as 3-position coordinates in a constraint space as planar coordinates describe positions on a plane. Constraint space coordinates are invariant under the rotation of the constraint space in 4-space in the same sense planar coordinates are independent of the orientation of the plane in 3-space. The concept has been extended to give constraint space coordinates $\pi_i$ and $\alpha_i$ in momentum and Bargmann 4-spaces. Integral transforms relating functions between the $\xi_i$, $\pi_i$ and $\alpha_i$ constraint spaces have been developed.

\end{document}